\documentclass[a4paper,12pt]{article} 
\usepackage{amssymb,amsmath,eepic,epsfig,psfrag}
\def\d{\mathrm{d}}
\def\D{\mathcal{D}}

\def\r{\mathbf{r}}
\def\R{\mathbf{R}}
\def\q{\mathbf{q}}
\def\ch{\mathrm{ch}}
\begin{document}
\title{\Large \bf Interfacial fluctuations near the critical filling 
transition}
\author{A.Bednorz and M.Napi\'orkowski\\
Instytut Fizyki Teoretycznej, Uniwersytet Warszawski,\\
00-681 Warszawa, Ho\.za 69, Poland} 
\maketitle 
\begin{abstract}
We propose a method to describe the short-distance behavior of an 
interface fluctuating in the presence of the wedge-shaped substrate 
near the critical filling transition. Two different length scales 
determined by the average height of the interface at the wedge center  
can be identified. On one length scale the one-dimensional approximation 
of Parry et al. \cite{Parry} which allows to find the interfacial critical
exponents is extracted from the full description. On the other scale 
the short-distance fluctuations are analyzed by the mean-field theory. \\

\end{abstract} 
\newpage

\centerline{\bf {I. Introduction}}
\renewcommand{\theequation}{1.\arabic{equation}} 
\setcounter{equation}{0}
\vspace*{0.5cm}

The analysis of physical systems usually involves some reductions in 
the description of the state of the systems. This is also the case for 
inhomogeneous systems consisting of two coexisting bulk phases 
separated by an interface fluctuating in the presence of a substrate. 
Certain properties of such systems, e.g. those related to the adsorption 
phenomena, can be conveniently described with the help of a single 
mesoscopic 
variable which is the distance of the interface from the substrate. \\
In this paper we consider such a system. The substrate has the 
form of an infinite wedge extending along the $y$-direction with the 
opening angle $2\varphi$, see Fig.1. The quasi-bulk phase adsorbed on the  
substrate is denoted as the $\beta$-phase while the phase far above the 
substrate is denoted as the $\alpha$-phase. The shape of the 
substrate is given by $z=|x|\,\cot\varphi$ and  $\ell(x,y)$ describes 
the distance of the $\alpha$-$\beta$ interface from this substrate. 
Recently, it was pointed out [1-7] that the above system may undergo the 
critical transition in which the position of the central part of the 
interface (above 
the edge of the wedge) moves to infinity while the asymptotic parts of 
the interface corresponding to $|x| \rightarrow \infty$ remain close 
to the substrate. This interfacial transition is called the filling 
transition to distinguish it from the wetting transition taking place on 
planar substrates \cite{Dietrich,Schick}. Thermodynamically the 
filling transition point is located at the bulk $\alpha$-$\beta$ 
coexistence and the filling temperature (which depends on the wedge 
opening angle $\varphi$) is denoted as  $T_{\varphi}$;   
$T_{\varphi} <T_{w}$, where $T_{w}$ is the wetting temperature on the 
planar substrate. \\ 
In their recent paper Parry et al. \cite{Parry} used the transfer-matrix 
method to  evaluate - among others - the values of the critical indices 
associated with the interfacial behavior near the filling transition. 
For this purpose the next step in the reduction of the description was 
made. The two-dimensional interface $\ell(x,y)$ was replaced by the 
one-dimensional mid-point line $\ell(y)\equiv\ell(0,y)$ (see Fig.1) for 
which the appropriate Hamiltonian was proposed. \\ 
If, however, one is interested in the full two-dimensional structure of 
the fluctuating interface near the critical filling transition, i.e. 
also in the short-distance behavior which is not included in the reduced 
description then - at least in principle -  one has to go 
beyond the mean-field analysis. In this paper we propose how to describe 
the two-dimensional interface close to the filling transition in the system 
with short-ranged forces \cite{RDN}. We expect that the geometry-dependent 
effects are important for short distances. Since the mid-point height 
does not vary too much on the short length-scale the idea is to fix this 
height at some arbitrarily chosen point and to assume the mean-field 
profile of the interface in the vicinity of this chosen point (along the 
$x$-direction). 
Then one uses the mean-field  approximation to describe the "relative" 
fluctuations around the fixed point. The two-point height distribution 
function for neighboring points consists of two parts: 
the one-point distribution corresponding to one of the points (or the 
average height of them) and the conditional probability distribution in 
the form of a Gaussian with position-dependent dispersion. Such quantity 
does not diverge at the filling transition and  may turn out useful when 
some geometry-dependent observables are considered. The mean-field 
description becomes then legitimate because by fixing the position of 
the interface and looking at the conditional distribution one forces the 
local fluctuations to be small and so one insists that the system is locally 
outside the critical region. \\

\centerline{\bf {II. The mean-field description}}
\renewcommand{\theequation}{2.\arabic{equation}} 
\setcounter{equation}{0}
\vspace{0.5cm}

\begin{figure}
\begin{center}
\begin{picture}(300,200)
\put(100,0){\vector(1,0){200}}
\put(110,150){\spline(0,1)(45,-3)(90,-5)(135,-2)(180,1)}
\put(10,50){\spline(0,1)(50,50)(100,101)}
\put(10,50){\spline(0,1)(-20,10)(-30,14)}
\put(110,150){\spline(0,1)(-20,10)(-30,16)}
\put(-20,65){\spline(0,-1)(35,35)(100,101)}
\put(190,50){\spline(0,-2)(35,35)(100,101)}
\put(190,50){\spline(0,-2)(20,10)(30,16)}
\put(290,150){\spline(0,1)(20,10)(30,14)}
\put(220,65){\spline(0,1)(50,50)(100,99)}
\put(200,100){\vector(1,1){70}}

\put(100,50){\spline(0,0)(-10,-2)(-20,-3)(-30,-2)(-40,0)(-50,2)(-60,3)(-70,2)
(-80,0)(-90,1)}
\put(100,0){\vector(0,1){50}}

\put(160,30){\vector(0,1){20}}
\put(100,50){\spline(0,0)(10,0)(20,5)(30,0)(40,-5)(50,-3)(60,0)}
\put(160,50){\spline(0,0)(10,-5)(20,-3)(30,-2)}
\put(275,175){$y$}
\put(305,0){$x$}
\put(163,52){$\ell(x,y)$}
\put(100,55){$\ell(y)$}
\put(100,0){\spline(-40,20)(-25,32)(0,40)}
\put(85,20){$\varphi$}
\put(120,30){$\beta$}
\put(120,70){$\alpha$}

\path(-24,62)(76,162)
\path(100,0)(200,100)
\put(100,0){\line(-2,1){124}}
\put(100,0){\line(2,1){124}}
\put(224,62){\line(1,1){100}}
\put(200,100){\line(-2,1){124}}
\put(200,100){\line(2,1){124}}
\end{picture}
\end{center}
\caption{The wedge geometry and the fluctuating $\alpha-\beta$ 
interface}\label{kli}
\end{figure}
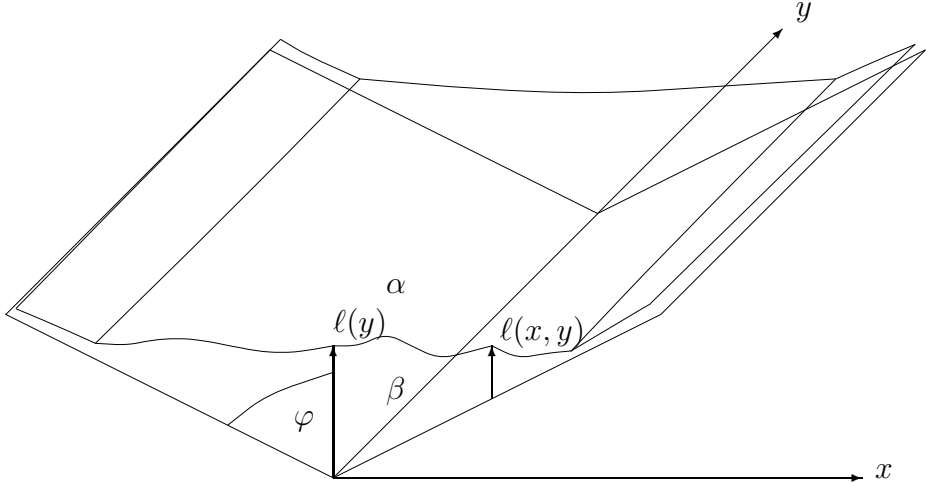

The interfacial Hamiltonian in the case of a very open wedge 
($\cot\varphi \ll 1$) has the standard form [6,10] 
\begin{equation}
\begin{split}
&H[\ell]=\int\d x\int\d y\left[\Sigma(\nabla(\ell+\alpha|x|)^2/2
+\omega(\ell)-\omega(\ell_{\infty})\right]\\&=
\int\d x\int\d y\left[\Sigma(\nabla\ell)^2/2
+\omega(\ell)-\omega(\ell_{\infty})\right]
-2\Sigma\alpha\int\d y[\ell(0,y)-\ell_{\infty}]\,,
\end{split}
\end{equation}
where $\ell(x,y)$ (Fig.1) denotes the width of the adsorbed $\beta$-like 
layer measured in the vertical direction and $\Sigma$ 
is the $\alpha$-$\beta$ interfacial tension. $\omega(\ell)$ denotes the 
interfacial pinning potential corresponding to the critical wetting 
in the planar case. For short-range forces considered in this paper it 
has the following form [6-11]
\begin{equation} 
\omega(\ell)=-W\,t\,\exp(-\ell/\xi) \,+\,U\,\exp(-2\ell/\xi) \,\,,
\label{ome}
\end{equation} 
where $\xi$ is the bulk correlation length (in the $\beta$-phase), 
$U$ and $W$ are positive constants. (We use the convention in which the 
factor $1/k_{B}T$ is included into the Hamiltonian.) 
The parameter $t$ denotes the  
dimensionless deviation from the wetting temperature for the 
planar substrate, i.e. $t > 0$ for $T<T_{w}$ and $t=0$ for $T=T_{w}$. 
$\ell_{\infty}$ is the equilibrium width of the adsorbed layer on 
the planar substrate which minimizes the potential $\omega(\ell)$: 
$\exp(-\ell_{\infty}/\xi) = \frac{W}{2\,U}t$. Because 
the wedge is very open we put $\sin\varphi =1$ and 
$\cot\varphi = \cos\varphi = \alpha$. \\ 

The mean-field profile $\bar{\ell}(x)$ varies only in the $x$ direction. 
It satisfies the Euler-Lagrange equation \cite{RDN} 
\begin{equation} 
\Sigma\bar{\ell}''(x) = \omega'(\bar{\ell}) \label{eula}
\end{equation}
and the  boundary conditions: $\bar{\ell}(\pm\infty)=\ell_{\infty}$, 
$\bar{\ell}'(0_\pm)=\mp\alpha$. 
The solution of Eq.(2.3) is 
\begin{equation}
x(\bar{\ell})=\pm\int_{\bar{\ell}}^{\ell_0}\frac{\d\ell}
{\sqrt{2(\omega(\ell)-\omega(\ell_{\infty}))/\Sigma}}\,\,,
\label{mfp}
\end{equation}
where the width of the mean-field profile at the center of the wedge 
$\ell_0=\bar{\ell}(0)$ satisfies 
$\omega(\ell_0)- \omega(\ell_{\infty})=\Sigma\alpha^2/2$. 
With the help of the Young equation one can relate $\omega(\ell_{\infty})$ to 
the contact angle $\Theta$ on the planar substrate:  
$-\omega(\ell_{\infty})=\Sigma\Theta^2/2$. From this we see that 
$\omega(\ell_0) =\Sigma(\alpha^2-\Theta^2)/2$ and the filling transition 
($\ell_{0}\rightarrow \infty$, $\ell_{\infty}$ -  finite) takes place when 
$\Theta(T=T_{\varphi})=\alpha$. \\ 
For small deviations $\delta\ell(x,y)=\ell(x,y)-\bar{\ell}(x)$ from the 
mean field profile $\bar{\ell}(x)$ the fluctuation Hamiltonian 
$H_{fl}[\delta\ell]=H[\bar{\ell}+\delta\ell]-H[\bar{\ell}]$ 
is bilinear in $\delta\ell$  
\begin{equation}
H_{fl}[\delta \ell]=\int \d x \int \d y\,\frac{1}{2}\left[
\Sigma(\nabla\delta\ell)^2+\omega''(\bar{\ell})(\delta\ell)^2\right]\,\,.
\end{equation}
The important feature of the critical filling transition is the 
existence of the \emph{translational mode}, i.e. the fluctuation of the 
interface which  
requires  very small energy (decreasing to $0$ at the filling point). This 
fluctuation takes the form $\delta\ell(x)=\epsilon|\bar{\ell}'(x)|$ and the 
corresponding energetical cost is 
$H_{fl}[\delta\ell] =\epsilon^2\alpha \omega'(\ell_0)$; it decreases to 
$0$ for $\ell_{0}\rightarrow \infty$.  \\ 

The corresponding differential equation for correlation function 
$G(\r,\r')=\langle\delta\ell(\r)\delta\ell(\r')\rangle$ has in the 
mean-field approximation the following form 
\begin{equation}
\left[-\Sigma\Delta_\r+\omega''(\bar{\ell})\right]G(\r,\r')=
\delta(\r-\r')\,\,.
\label{corel}
\end{equation}

However, the mean-field description fails in case of the critical  
filling transition for short-ranged forces because Eq.(2.6) implies 
strong anisotropy of fluctuations of the interface. 
The fluctuations along the wedge diverge much faster than across the wedge. 
The latter are bounded by the geometry of the substrate. As shown in 
\cite{Parry}, the mean-field predictions are valid only for power-law forces 
of the type $\omega(\ell)\sim\ell^{-p}$ for $p<4$. \\

\centerline{\bf {III. The reduction of the order parameter}}
\renewcommand{\theequation}{3.\arabic{equation}}
\setcounter{equation}{0}
\vspace{0.5cm}

The \emph{effective} way to analyze the critical fluctuations of 
$\ell(x,y)$ near the filling transition point is to reduce the interfacial 
description by looking only at the mid-point height $\ell(y)=\ell(0,y)$ [10]. 
In order to derive the corresponding Hamiltonian we proceed as follows: we 
minimize the Hamiltonian in Eq.(2.1) similarly as in the mean-field method 
but now with the constraint $\ell(0,y)=\ell(y)$  imposed  independently at 
each $y$ \cite{FJP}. From the corresponding Euler-Lagrange equation one 
obtains 
\begin{equation}
x(\ell,y)=\pm\int_\ell^{\ell(y)}\frac{\d\ell_1}
{\sqrt{2(\omega(\ell_1)-\omega(\ell_{\infty}))/\Sigma}}\,\,. 
\label{loc}
\end{equation}
As the result the one-dimensional Hamiltonian $H_1[\ell(y)]=H[\ell(x,y)]$
valid for configurations given in Eq.(\ref{loc}) takes the form 
\begin{equation}
\begin{split}
&H_1[\ell(y)]=\int\d y
\left\{\frac{\Sigma(\ell'(y))^2\int_{\ell_{\infty}}^{\ell(y)}\d\ell_1 
\sqrt{\Sigma(\omega(\ell_1)-\omega(\ell_{\infty}))/2}}
{\omega(\ell(y))- \omega(\ell_{\infty})}\right. 
\\&\left.+2\int_{\ell_{\infty}}^{\ell(y)}\d\ell_1\left[\sqrt{2\Sigma(
\omega(\ell_1)-\omega(\ell_{\infty}))}-\alpha\Sigma\right]\right\}. 
\end{split}
\end{equation}
For short-range forces, see Eq.(\ref{ome}), the above Hamiltonian can be 
explicitly evaluated 
\begin{equation}
\begin{split}
&H_1[\ell]=\int\d y\left\{\frac{\Sigma(\tilde{\ell}'(y))^2}
{\Theta}\frac{\tilde{\ell}-\xi(1-\exp{(-\tilde{\ell}/\xi)})}
{[1-\exp{(-\tilde{\ell}/\xi)}]^2} \right.\\&
\left.+2\Sigma[(\Theta-\alpha)\tilde{\ell} \,-
\Theta\xi(1-\exp{(-\tilde{\ell}/\xi)})]\right\}\,\,,
\end{split}
\end{equation}
where $\tilde{\ell}(y)=\ell(y)-\ell_{\infty}$. For temperatures close 
to the filling transition one has $\ell\gg\xi$ and $\ell\gg\ell_{\infty}$ 
and Eq.(3.3) reduces to the one-dimensional Hamiltonian proposed 
phenomenologically in \cite{Parry} 
\begin{equation}
H_1[\ell(y)]\approx\int\d y\left
[\frac{\Sigma\ell(y)}{\Theta}(\ell'(y))^2+2\Sigma(\Theta-\alpha)\ell(y)
\right].
\end{equation}
The above Hamiltonian has relatively simple structure and is easy to 
renormalize. After introducing the rescaled variables 
$L$ and $Y$ 
\begin{equation}
\Theta y=(2\Sigma)^{-1/2}(\Theta/\alpha-1)^{-3/4}Y\,\,,\,\,
\quad\ell=(2\Sigma)^{-1/2}(\Theta/\alpha-1)^{-1/4}L 
\end{equation}
it takes the form
\begin{equation}
H_1[L(Y)]=\int\d Y\left[\frac{L(Y)}{2}(L'(Y))^2+L(Y)\right]\,,
\label{1dH}
\end{equation}
which is free from parameters. Accordingly, the critical behavior of 
the mean mid-point height $\langle\ell(y)\rangle$ and the correlation 
length $\xi_{y}$ follow directly from the above rescaling: 
$\langle\ell(y)\rangle\sim(\Theta-\alpha)^{-1/4}$ 
and  $\xi_y\sim(\Theta-\alpha)^{-3/4}$. The values of the critical 
indices agree with those obtained in [10]. \\

The one-dimensional model described by the Hamiltonian in Eq.(\ref{1dH}) 
can be solved via the transfer-matrix method \cite{Parry, Burk}. 
However, in this method the presence of the factor $L(Y)$ in front of 
$(L'(Y))^2$ is the source of ambiguity while discretizing the problem 
and defining the measure which is then used to evaluate the relevant 
propagator \cite{BN}. In order to avoid such problems it is convenient 
to introduce the new variable $\eta\equiv 2L^{3/2}/3$ which "absorbs" the 
dangerous factor $L(Y)$ in front of $(L'(Y))^2$. Then the Hamiltonian 
takes the form 
\begin{equation}H_1[\eta(Y)]=\int\d Y\left[(\eta'(Y))^2/2+(3\eta/2)^{2/3}
\right]\,.
\end{equation}
The corresponding propagator 
\begin{equation}
V(\eta_2,\eta_1,Y)=\int\D\eta\exp(-H_1[\eta])|_{\eta(0)=\eta_1}^
{\eta(Y)=\eta_2}
\end{equation} can be evaluated by solving - within the transfer matrix 
approach \cite{Burk} - the following equation 
\begin{equation}
\frac{\partial V}{\partial Y}=\frac{\partial^2V}{2\partial\eta_2^2}-
(2\eta_2/3)^{2/3}V.
\end{equation}
This equation must be supplemented by the appropriate boundary 
condition for $\eta_2=0$. The general form of such a condition 
\begin{displaymath}
\partial_{\eta_2}\ln V|_{\eta_2=0}=a_t 
\end{displaymath}
is similar to that found in \cite{Burk} for 2D wetting. In the present 
case the parameter $\eta$ is $t$-dependent so one expects the 
$t$-independent boundary condition 
\begin{displaymath}
\partial_{\ell_2^{3/2}}\ln V=a=(2\Sigma)^{3/4}
(\Theta/\alpha-1)^{3/8}a_t \,.
\end{displaymath}
For $a<0$ the edge effects become dominant and no filling is observed. Thus 
we assume  $a>0$ ($a^{-2/3}$ is the range of the influence of the edge 
effects), so  $a_t \sim(\Theta-\alpha)^{-3/8}$ in the critical 
region. The appropriate boundary condition is then $V(0,\eta_1,Y)=0$. \\ 
The propagator $V(\eta_2,\eta_1,Y)$ can be expressed by 
normalized eigenfunctions $\psi_n(\eta)$ and eigenvalues $E_n$ of the 
equation
\begin{equation}
E_n\psi_n=-\frac{\partial^2\psi_n}{2\partial\eta^2}+(3\eta/2)^{2/3}\psi_n\,.
\end{equation}
Then  
\begin{equation}
V(\eta_2,\eta_1,Y)=\sum_n\psi_n(\eta_1)\psi_n(\eta_2)
e^{-E_nY}\,. 
\end{equation}
The probability distribution of the mid-point height is given by 
$\psi_0^2(\eta)$ and other quantities can be expressed by the appropriate 
combinations of eigenfunctions. \\

\centerline{\bf {IV. The short-distance correlation function}}
\renewcommand{\theequation}{4.\arabic{equation}}
\setcounter{equation}{0}
\vspace{0.5cm}

Obviously the above one-dimensional approximation cannot describe the full 
two-dimensional structure of the interface. However, there are two 
different length-scales in this problem. The one-dimensional character 
of the filling transition is seen on scales $\alpha y\sim\Sigma\ell^3$ 
while the two-dimensional structure becomes important when 
$\alpha y\sim\ell$. In the critical region these two scales are well 
separated. \\  Therefore, in order to analyze the short-distance behavior 
one can introduce \emph{the conditional correlation function}. This is 
done in the following way. We assume that for certain $y_0$ 
(for convenience we set $y_0=0$) the interface profile $\ell(x,y)$ is 
constrained: $\ell(x,y_0=0)=\bar{\ell}(x)$, where $\bar{\ell}(x)$ 
is described by Eq.(\ref{mfp}) but with {\it {given}} $\ell_0$. The full 
Hamiltonian is then expanded in the Taylor series in the variable 
$\phi(x,y)=\ell(x,y)-\ell(x,0)$ up to $\phi^2$ terms. In this way one 
obtains (up to the constant term) 
\begin{equation}
\begin{split}&H[\phi]=-2\int\d y\left[\alpha\Sigma-
\sqrt{2\Sigma(\omega(\ell_0)-\omega(\ell_{\infty}))}
\right]\phi(0,y)\\&+\frac{1}{2}\int\d x\int\d y\left[\Sigma(\nabla\phi)^2
+\omega''(\bar{\ell}(x))\phi^2\right].
\end{split}
\label{hphi}
\end{equation}
The first term on the rhs of Eq.(4.1) is very small in the critical region 
(i.e. for $\ell_{0} \rightarrow \infty$ and $\Theta \approx \alpha$); it is 
given by 
$\sim 2\Sigma(\Theta-\alpha)$. Thus for short distances one keeps only the 
second term. The resulting structure of the Hamiltonian implies the following 
differential equation for the conditional correlation function 
\begin{displaymath}
G_{\ell_0}(\r,\r')=\langle\phi(\r)\phi(\r')\rangle|_{\ell(x,0)=\bar{\ell}(x)}
\end{displaymath}
\begin{equation}
\left[-\Sigma\Delta_\r+\omega''(\bar{\ell})\right]G_{\ell_{0}}(\r,\r')
=\delta(\r-\r')\,.
\end{equation}
Similarly as in Eq.(3.11) the conditional correlation function can be 
expressed 
by the normalized eigenfunctions $\psi_\q$ and eigenvalues $E_\q$ of the 
operator $[-\Sigma\Delta+\omega''(\bar{\ell})]$
\begin{displaymath}
G_{\ell_{0}}(\r,\r')=\sum_\q\frac{\psi_\q(\r)\psi_\q(\r')}{E_\q}\,.
\end{displaymath}
In this approach one has to analyze carefully the contribution from the 
eigenvalues tending to $0$. One expects that the eigenfunctions with the 
lowest eigenvalues will have their structure similar to 
$\psi_0=|\bar{\ell}'(x)|$ which itself corresponds to the translational mode 
(although it does not satisfy the appropriate boundary condition in the 
present case). Thus we introduce the new variables $\psi_\q=\varphi_\q\psi_0$ 
and the equation for $\varphi_\q$ has the form 
\begin{equation}
\left[E_\q+\Sigma\Delta\right]\varphi_\q=-2\left[\sqrt{2\Sigma(\omega(\bar
{\ell})-\omega(\ell_{\infty}))}\right]'\partial_x\varphi_\q\,\,. 
\end{equation}
The expression on the rhs of the above equation is close to $0$ for 
$\alpha|x|<\ell_0$ and for $\alpha|x|$ approaching $\ell_{0}$ it quickly 
becomes equal to 
$-2\sqrt{\Sigma\omega''(\ell_\infty)} \partial_x\varphi_\q$. We are 
interested only in the long-wave fluctuations such that  
$E_\q\sim\Sigma(\alpha/\ell_0)^2$. If all terms in the above equation are 
to be of the same order of magnitude for $\alpha|x|>\ell_0$ then one 
should have 
$\partial_x\ln\varphi_\q\sim E_\q\xi_\pi/\Sigma\sim\xi_{\pi}
\alpha^2/\ell_0^2$,  
where $\xi_\pi=(\omega''(\ell_\infty)/\Sigma )^{-1/2}$ 
is the correlation length for the planar case. Note that $\xi_{\pi}$ which 
diverges at the critical wetting on the planar substrate 
remains finite at the critical filling transition. \\ 
The above considerations lead to the following equation for the conditional 
correlation function for $\alpha|x|<\ell_0$, i.e for the central "free" 
part of the interface 
\begin{eqnarray}
-\Sigma\Delta_\r G_{\ell_0}(\r,\r')&=&\delta(\r_2-\r_1)\\\partial_x
G_{\ell_0}(\r,\r')|_{|x|=\ell_0/\alpha}&=&0.
\end{eqnarray}
One also needs the boundary condition for $y'\rightarrow\infty$. 
Since there exists the long-range order on the scale considered now, i.e. 
for $\alpha y \sim \ell_{0}$ one should not expect 
$G_{\ell_0}(\r_1,\r_2)\stackrel{\r_2\rightarrow\infty}{\longrightarrow}0$. 
Instead we assume 
$G_{\ell_0}(\r_1,\r_2)\stackrel{\r_2\rightarrow \infty}{\longrightarrow}
f(\r_1)<\infty$, i.e. $G$ remains finite. Using the standard methods of 
conformal transformations (see Appendix) one obtains the following 
solution of Eqs.(4.4, 4.5) 
\begin{equation}
\begin{split}
G_{\ell_0}(\r_1,\r_2)&=-\frac{1}{4\Sigma\pi}\left[\ln\left(
e^{(Y_1-Y_2)\pi/2}+e^{(Y_2-Y_1)\pi/2}
-2\cos(X_1-X_2)\pi/2\right)\right.\\
&+\ln\left(e^{(Y_1-Y_2)\pi/2}+e^{(Y_2-Y_1)\pi/2}
+2\cos(X_1+X_2)\pi/2\right)\\
&-\ln\left(e^{(Y_1+Y_2)\pi}+1
-2e^{(Y_1+Y_2)\pi/2}\cos(X_1-X_2)\pi/2\right)\\
&\left.-\ln\left(e^{(Y_1+Y_2)\pi}+1
+2e^{(Y_1+Y_2)\pi/2}\cos(X_1+X_2)\pi/2\right)+\pi(Y_1+Y_2)\right]
\end{split}\label{corl}
\end{equation}
where $\R_{i}=\alpha\r_{i}/\ell_0$, $i=1,2$. We note that for  
$\r_2\rightarrow \infty$ one has $G_{\ell_0}(\r_1,\r_2) \rightarrow  
\alpha y_1/\Sigma \ell_{0}$. \\

\centerline{\bf {V. The short-distance dispersion }}
\renewcommand{\theequation}{5.\arabic{equation}}
\setcounter{equation}{0}
\vspace{0.5cm}

\begin{figure}

\epsfxsize=6.5cm
\psfrag{sig}{$\sigma$}
\psfrag{x2}{$\scriptstyle X_2\;$}
\psfrag{x1}{$\scriptstyle X_1\;$}
\epsffile{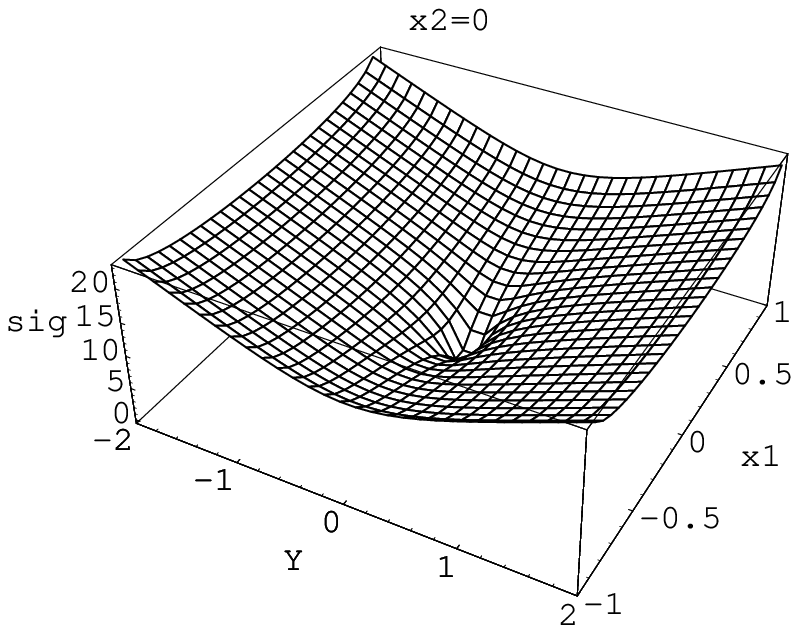}
\epsfxsize=6.5cm
\epsffile{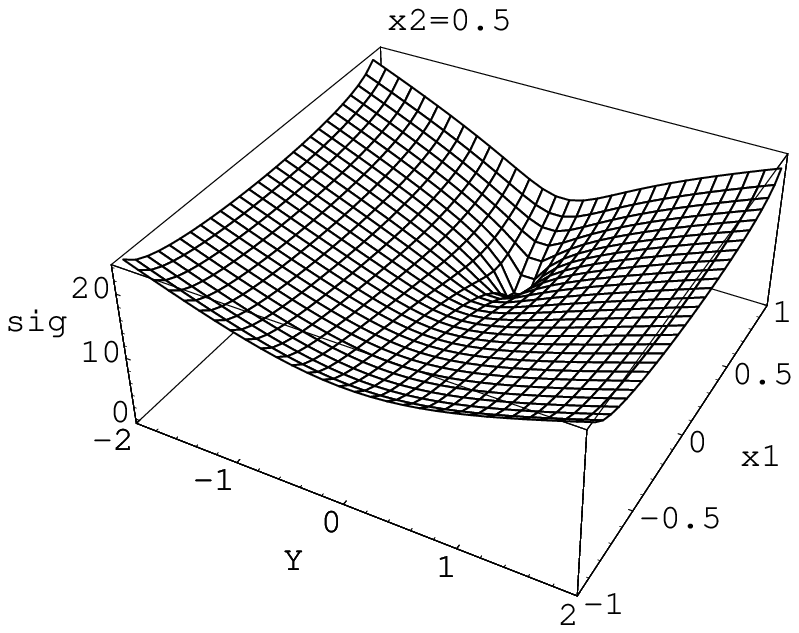}
\caption{The dimensionless dispersion $\sigma$ as function of $X_1$ and $Y$
for $X_2=0$ and $0.5$, respectively.}\label{sii}
\end{figure}

For short distances the two-point $\ell$-distribution function has 
the form 
\begin{equation}
p(\ell_1,\r_1;\ell_2,\r_2)\approx p(\ell_0)
\frac{\exp\left(-(\ell_2-\ell_1)^2/
2\sigma(\r_1,\r_2,\ell_{0})\right)}{[2\pi\sigma(\r_1,\r_2,\ell_{0})]^{1/2}}
\end{equation}
where $\ell_0$ is the height of the interface above the edge of the wedge 
\begin{displaymath}
\ell_{0}=(\ell_1+\ell_2+\alpha(|x_1|+|x_2|))/2\approx\ell_1+\alpha|x_1|
\approx\ell_2+\alpha|x_2|\,\,. 
\end{displaymath}
We use the conditional correlation function $G_{\ell_0}$ to obtain the 
expression for the dispersion $\sigma$
\begin{equation}
\sigma=\langle(\ell_2-\ell_1)^2\rangle=G_{\ell_0}(\r_1,\r_1)
-2G_{\ell_0}(\r_1,\r_2)+G_{\ell_0}(\r_2,\r_2)\,\,.
\end{equation}
The standard problem which one encounters at this point is that 
$G_{\ell_0}(\r_1,\r_2)$ diverges for $\r_2 \rightarrow \r_1$ \cite{HWD,BGZ}. 
This divergence 
can be removed by regularizing the function $G_{\ell_0}(\r_1,\r_2)$, e.g.  
by adding to the Hamiltonian given in Eq.(\ref{hphi}) the term 
$a^2(\Delta\phi)^2/2$, where $a$ is a dimensionless parameter.  This 
procedure yields the following equation for the regularized function 
$G_{\ell_0}^{(a)}(\r_1,\r_2)$ 
\begin{displaymath}
\left[a^2{\Delta^2_{\r_{1}}}-\Sigma\Delta{_{\r_{1}}}\right]G_{\ell_0}^{(a)}=
\delta(\r_2-\r_1)\,.
\end{displaymath}
For small $a$ the solution of the above equation has the form 
\begin{displaymath}
G_{\ell_0}^{(a)}(\r_1,\r_2)=G_{\ell_0}(\r_1,\r_2)-
K_0(\Sigma^{1/2}|\r_2-\r_1|/a)/2\pi\Sigma\,,
\end{displaymath}
where $K_0$ is the modified Bessel function. 
In this way the short-distance divergence is removed and one has 
\begin{displaymath}
G_{\ell_0}^{(a)}(\r_1,\r_1)=\lim_{\r_2\rightarrow\r_1}\left[
G_{\ell_0}(\r_1,\r_2)+\frac{\gamma+\ln(\Sigma^{1/2}|\r_2-\r_1|/2a)}
{2\pi\Sigma}\right]\,,
\end{displaymath}
where $\gamma$ is the Euler constant. Now the expression for the dispersion 
$\sigma(\r_1,\r_2,\ell_0)$, Eq.(5.2) can be written down explicitly. We are 
interested in the situation in which the constraint 
affects only the mean height of the interface and thus we consider the case 
$y_1,y_2\gg\ell_0/\alpha$ and $|\r_1-\r_2|\gg a$. Then 
\begin{equation}
\begin{split}
&\sigma(\r_1,\r_2,\ell_0)=\frac{1}{2\Sigma\pi}
\left\{2\ln\left(
\ell_0\Sigma^{1/2}/a\pi\alpha\right)+2\gamma
-\ln\left(\cos (X_1\pi/2))
\right)\right.\\&-\ln\left(\cos (X_2\pi/2)\right)
+\ln\left[\ch((Y_1-Y_2)\pi/2)-\cos((X_1-X_2)\pi/2)
\right]\\
&\left.+\ln\left[\ch((Y_1-Y_2)\pi/2)+ \cos((X_1+X_2)\pi/2)
\right]\right\}\,. 
\end{split}
\end{equation}
The behavior of $\sigma$ is shown on Fig.\ref{sii}.
We see that in this limit $\sigma(\r_1,\r_2,\ell_0)$ depends - in addition 
to $X_1$ and $X_2$ - only on the distance $Y = Y_2 - Y_1$. For 
fixed values of  $X_1$ and $X_2$ it is an increasing function of $|Y|$, 
see Fig.2.  Thus the relative fluctuations of the interface position at 
points distant along the edge of the wedge become large. \\ 
It is interesting to observe that for  $|y_1-y_2|\gg\ell_0/\alpha$ one gets 
\begin{displaymath}
\sigma(\r_1,\r_2,\ell_0)\approx|y_1-y_2|\alpha(\Sigma\ell_0)^{-1}/2\,.
\end{displaymath}
This result agrees with the prediction of the one-dimensional model valid 
on the scale where $\Sigma\ell_0^3\gg\alpha|y_1-y_2|$. It can be derived 
with the help of Eq.(3.9). Thus the results obtained via the conditional 
correlation function in Chapters IV and V  are consistent with those stemming 
from the transfer-matrix analysis of the 1D model in Chapter III. \\

\centerline{\bf {VI. Conclusions }}
\vspace{0.5cm}

The reduced description of the interface fluctuating in the presence of 
the wedge-shaped substrate is derived in an explicit way. This reduced 
description is based on the one-dimensional Hamiltonian \cite{Parry} and 
the presented derivation of this Hamiltonian makes clear use of the physical 
assumptions behind it. Although the one-dimensional Hamiltonian allows one to 
find the relevant critical exponents it cannot describe the full 
two-dimensional structure of the interface. We have proposed the method of 
supplementing this  one-dimensional picture by the local two-dimensional 
constrained fluctuations which can be analyzed within the mean-field theory 
and described by the conditional correlation function.  These 
fluctuations are not divergent at the filling transition. 
The proposed method can be used to calculate the geometry-dependent 
observables. Moreover, it predicts the behavior of the dispersion of the 
conditional correlation function which agrees with the predictions of the 
one-dimensional model.\\

{\bf Acknowledgment}
The authors gratefully acknowledge the discussions with H.W.Diehl, 
S.Dietrich, and A.Parry, and the support by the Foundation for German-Polish 
Collaboration under Grant. No. 3269/97/LN. \\

\centerline{\bf {Appendix }}
\renewcommand{\theequation}{A.\arabic{equation}}
\setcounter{equation}{0}
\vspace{0.5cm}

In this Appendix we sketch the consecutive steps of the method of 
conformal transformations which lead to the solution of Eqs.(4.4,4.5). \\ 
After introducing the complex variables  $\r_{1,2}=(x_{1,2},y_{1,2})$, 
$z_{1,2}=x_{1,2}+iy_{1,2}$ Eq.(4.4) can be rewritten as 
\begin{equation} 
-4\Sigma\partial_{z_1}\partial_{\bar{z}_1}G_{\ell_0}(z_1,z_2)
=\delta(z_1-z_2).
\label{kon}
\end{equation}
together with the  boundary condition (Eq.(4.5))
\begin{equation}
i[\d z_c\partial_z-\d\bar{z}_c \partial_{\bar{z}}]\,G_{\ell_0}(z,z_2)=0\,,
\label{brz}
\end{equation} 
where $z_c$ denotes the contour on which the boundary condition is given. 
Eqs.(A.1, A.2) are invariant with respect to the conformal transformations. 
The solution of Eq.(A.1) valid for the whole plane has the form 
\begin{equation}
G_\infty(z_1,z_2)=-\frac{1}{4\Sigma\pi}[\log(z_1-z_2)+\log
(\bar{z}_1-\bar{z}_2)].
\end{equation}
The solution valid for the semi-plane  $\Re z>0$ with the Neumann condition 
on  $\Re z=0$ is found with the help of the method of images and has the 
form 
\begin{equation}
G_{\Re z>0}(z_1,z_2)=G_\infty(z_1,z_2)+G_\infty(z_1,-\bar{z}_2).
\end{equation}
After introducing the conformal transformation  
$Z\mapsto e^{-i\pi Z/2}$ for dimensionless variables 
$Z=z\alpha/\ell\,,\mathbf R=\r\alpha/\ell$ one obtains 
\begin{equation}
G_{(\ell)}(Z_1,Z_2)=G_{\Re Z >0}(e^{i\pi Z_1/2}, 
e^{i\pi Z_2/2})
\end{equation}
with the Neumann condition at $X=\pm 1$. 
Finally, taking into account the constraint imposed on the interface 
at $y=0$ and using the freedom to add to the rhs of Eq.(A.4) the solutions 
of the Laplace equation leads to the solution of Eq.(4.4) given in Eq.(4.6).

\newpage
\end{document}